\documentclass[aps,prl,twocolumn,groupedaddress,amsmath,amssymb,amsfonts,graphics]{revtex4}

\newcommand{\crit}[1]{ {#1}_{\text{c}} }
\newcommand{\cross}[1]{ {#1}_{\text{i}} }

\usepackage{graphicx}
\usepackage{bm}
\usepackage{bbold}

\begin{document}

\title{Data collapse in the critical region using finite-size scaling with subleading corrections}

\author{K.\ S.\ D.\ Beach}
\author{Ling Wang}
\author{Anders W.\ Sandvik}
\affiliation{Department of Physics, Boston University, 590 Commonwealth Avenue, Boston, MA 02215}

\date{May 8, 2005}

\begin{abstract}
We propose a treatment of the subleading corrections to finite-size scaling
that preserves the notion of data collapse.
This approach is used to extend and improve the usual Binder cumulant analysis.
As a demonstration, we present results for the two- and three-dimensional classical Ising models
and the two-dimensional, double-layer quantum antiferromagnet.
\end{abstract}

\maketitle

True phase transitions occur only in systems with an infinite number of degrees of freedom. 
Nevertheless, it is possible to study critical phenomena by simulating \emph{finite} systems of 
increasing size and extrapolating their behaviour to the thermodynamic limit. 
The standard analysis relies on Fisher's finite-size scaling (FSS) 
hypothesis~\cite{Fisher71,Fisher72}, which supposes
that near a critical point every thermodynamic property of the system scales as a universal function of 
the ratio $L/\xi$, where $L$ is the linear size of the system and $\xi$ is the bulk correlation length.

The basic ansatz is that any property with critical exponent $\kappa$
obeys the relation
$A(T,L) = L^{\kappa/\nu} f_{A}(L/\xi(t))$, which is valid 
for large $L$ and small reduced temperature
$t = (T - \crit{T})/\crit{T}$. Here, $\crit{T}$ is the critical temperature
of the transition, and $\nu$ is the critical exponent of
the correlation length $\xi \sim \lvert t \rvert^{-\nu}$. 
The function $f_{A}(x)$ depends only on the boundary conditions and the 
system geometry. Equivalently, one can write
\begin{equation} \label{EQ:FSSansatz}
A(T,L)L^{-\kappa/\nu} = g_{A}(tL^{1/\nu}).
\end{equation}

Equation~\eqref{EQ:FSSansatz} can be formally motivated~\cite{Brezin82,Barber83,Brezin85} 
by arguing that the diverging correlation length leaves the
system scale invariant~\cite{Stanley99}.
As $t\rightarrow 0$,
fluctuations of the dynamical variables begin to look the same on all 
length scales (much greater than the lattice spacing).
Consequently, any change in the system size $L$ can be compensated by a 
corresponding change in temperature $t$. The
two otherwise independent quantities become locked together in a
single composite variable $x = tL^{1/\nu}$.

Graphically, this effect is undersood in terms of
so-called \emph{data collapse}~\cite{Binder70_80,Ferron04}.
A plot of Eq.~\eqref{EQ:FSSansatz} in reduced variables
traces out the universal function $g_A(x)$.
Such a plot can be useful as a data analysis tool.
Given some set of measurements
$\{ (T_i,L_i,A_i) \}$, 
the critical behaviour
of the system
can be determined by constructing the coordinates
$x_i = L_i^{1/\nu}(T_i-\crit{T})/\crit{T}$ 
and
$y_i = A_iL_i^{-\kappa/\nu}$
and choosing the values of $\nu$, $\kappa$, and $\crit{T}$ such that
the graph points $\{(x_i,y_i)\}$ have the least amount of scatter~\cite{endnote1}.

The accuracy of such a fit depends not only on the quantity and quality
of the data; it also depends sensitively on whether the data 
are sufficiently deep in the scaling limit ($L\rightarrow \infty$ at fixed $x$).
What constitutes ``deep enough,'' however, is model dependent and difficult to determine.
For high-accuracy studies, it is best not to put complete trust in Eq.~\eqref{EQ:FSSansatz}.
Since true scaling is realized only asymptotically, data from all but the largest systems 
can rarely be made to fall onto a single curve. Worse, the data can sometimes
be shoehorned into collapse with \emph{incorrect} parameters.
For lattice sizes that are accessible to computer 
simulation, there are almost always significant, non-universal corrections to FSS.
To simply ignore them can lead to subtle systematic errors,
unreliable fits, and overly-optimistic error estimates.

In this Letter, we show how to incorporate subleading corrections
into the minimal-scatter optimization  described above.
We argue
that the correct approach is to make
small, size-dependent modifications to the prefactor 
\emph{and argument} of $g_A(x)$ in Eq.~\eqref{EQ:FSSansatz}:
\begin{equation} \label{EQ:FSSanstatz2}
A(T,L)L^{-\kappa/\nu} = \mathcal{N}(L)g_A(tL^{1/\nu}-\epsilon(L)).
\end{equation}
In other words, the features of $A$ are both
renormalized \emph{and shifted}.
In the thermodynamic limit, $\mathcal{N}(L) \rightarrow 1$ and $\epsilon(L) \rightarrow 0$;
we determine the precise asymptotic form of these functions by
renormalization group arguments. The key point
is that $\tilde{y} = A(T,L)L^{-\kappa/\nu}\mathcal{N}(L)^{-1}$ now behaves as a universal function
of $\tilde{x} = tL^{1/\nu} - \epsilon(L)$.

In the special case where $g_A(x)$ is bounded and monotonic,
$A(T,L)L^{-\kappa/\nu}$ becomes increasingly step-function-like
as $L\rightarrow \infty$. Thus, two such 
curves, plotted versus temperature for different linear sizes $L$ and $L'$, 
will intersect (except in unusual circumstances)
at a unique point in the vicinity of the critical temperature $\crit{T}$.
Binder's spin-distribution cumulants~\cite{Binder81} 
have this useful property
(as does, \emph{e.g.}, the spin stiffness~\cite{Wang05}).
In the Binder cumulant case, $\kappa = 0$ happens to be known \emph{a priori}.
We solve formally for the point of intersection $\cross{T}(L,L')$ under the assumption of Eq.~\eqref{EQ:FSSanstatz2}-like scaling and determine its location as a function of $L$ and $n = L'/L$. We prove that a sequence of intersection points ($n$ fixed, $L \rightarrow \infty$) converges to $\crit{T}$ faster than $L^{-1/\nu}$.
Fitting the $n$ dependence of a data set offers another way of estimating $\crit{T}$ and $\nu$.

\emph{Corrections to finite-size scaling}---FSS can be put on a (somewhat) rigourous basis 
by invoking the
renormalization group (RG)~\cite{Brezin82,Barber83,Brezin85}.
From the RG point of view, the system is characterized by a set of 
scaling fields $\{\zeta_i\}$. 
The singular part of the free energy behaves according to
\begin{equation} \label{EQ:freeenergy1}
f_{\text{s}}(\zeta_1, \zeta_2, \zeta_3, \ldots) 
= L^{-d}\mathcal{F}(\zeta_1 L^{y_1},\zeta_2 L^{y_2},\zeta_3 L^{y_3}, \ldots),
\end{equation}
where
$\mathcal{F}(x_1,x_2,x_3,\ldots)$ is a regular function of its arguments.
The $y_i$ are eigenvalues of the RG transformation and characterize
the flow of the fields under the rescaling $L \mapsto L/b$ with $b > 1$;
\emph{i.e.}, $\zeta_i \mapsto \zeta_i b^{y_i}$.

For concreteness, suppose that there is only one relevant scaling
field $\zeta_1= t$ with a scaling exponent $y_1 = 1/\nu > 0$.
Then $f_{\text{s}}$ admits a series expansion in the remaining fields
(with $\zeta_i=\zeta_{i,0}$ fixed):
\begin{equation} \label{EQ:freeenergy2}
f_{\text{s}} \!=\! L^{-d} \prod_{i=2}^{\infty} \sum_{n = 0}^\infty  
\frac{\zeta_{i,0}^nL^{-n\lvert y_i\rvert}}{n!}
\frac{\partial^{n}}{\partial x_i^{n}} 
\mathcal{F}(tL^{1/\nu}\!,x_2,x_3,\ldots)
\biggr\rvert_{x_i = 0}.
\end{equation}
It follows that any thermodynamic quantity generated from
Eq.~\eqref{EQ:freeenergy2} 
will have the form
\begin{equation} \label{EQ:FSSexpansion}
A(T,L) = L^{\kappa/\nu} g_A(tL^{1/\nu}) + L^{(\kappa-\phi)/\nu} p_A(tL^{1/\nu}) + \cdots
\end{equation}
where $\phi/\nu = \min(\lvert y_i \rvert)$.
The functions $g_A(x)$ and $p_A(x)$
are partial derivatives of $\mathcal{F}$.
Their asymptotic behaviour, $g_A(x) \sim \lvert x \rvert^{-\kappa}$ and 
$p_A(x) \sim \lvert x \rvert^{-\kappa+\phi}$ as $x \rightarrow \pm \infty$,
leads to the thermodynamic limit
\begin{equation}
A(T) = \lim_{L\rightarrow \infty} A(T,L) \sim \lvert t \rvert^\kappa \bigl( 1 + (\text{const}) \lvert t \rvert^\phi + \cdots \bigr).
\end{equation}

The functions $g_A(x)$ and $p_A(x)$ are well-behaved and admit series expansions
$g_A(x) = g_0 + g_1x + \frac{1}{2}g_2x^2 + \cdots$, \emph{etc.}
Identifying Eqs.~\eqref{EQ:FSSanstatz2} and \eqref{EQ:FSSexpansion}
up to $O(x^3)$, we find that
$\mathcal{N}(L) = 1 + C_1L^{-\phi/\nu}$
and $\epsilon(L) = C_2L^{-\phi/\nu}$,
where
\begin{equation}
C_1 = \frac{g_2p_0 - g_1p_1}{g_2g_0 - g_1^2}
\ \ \text{and} \ \
C_2 = \frac{g_0p_1 - g_1p_0}{g_2g_0 - g_1^2}.
\end{equation}

This analysis is flawed, however,  in that scaling holds only asymptotically.
The ``equalities'' in Eqs.~\eqref{EQ:freeenergy1} and \eqref{EQ:freeenergy2}
are subject to corrections analytic in $L^{-1}$. This means that,
strictly speaking, 
$\mathcal{N}(L) = ( 1 + C_1L^{-\phi/\nu} + \cdots)(1 + a_1L^{-1} + a_2L^{-2} + \cdots )$.
Over some range of $L$ values, this can be represented by 
$\mathcal{N}(L) = 1 + c L^{-\omega}$, where 
$\omega$ has some effective value close to that of the true $L \rightarrow \infty$ exponent,
$\min(\phi/\nu,1)$. 
Thus,
\begin{equation} \label{EQ:FSSsubleading}
A(T,L) = L^{\kappa/\nu}\bigl(1 + cL^{-\omega})g_A(tL^{1/\nu}-dL^{-\phi/\nu})
\end{equation}
is the appropriate generalization of Eq.~\eqref{EQ:FSSansatz}.
Most important, this scaling ansatz represents
an improvement over Eq.~\eqref{EQ:FSSexpansion}
in that data collapse can still be engineered by plotting
$AL^{-\kappa/\nu}(1 + cL^{-\omega})^{-1}$ versus $tL^{1/\nu}-dL^{-\phi/\nu}$,
with the addition of $\omega$, $\phi$, $c$, and $d$ as fitting parameters.

\emph{Intersection point analysis}---In the vicinity of a magnetic phase transition, expectation values of powers of the 
magnetization obey the scaling relation
$\langle \lvert M \rvert ^p \rangle_L = L^{-p\beta/\nu} g_{|\!M\!|^p}(tL^{1/\nu})$ for integer $p > 0$.
At large $x$, the scaling function $g_{|\!M\!|^p}(x)$
exhibits power-law behaviour with exponents 
$p\beta$ as $x\rightarrow -\infty$ and $p\tilde{\beta} < p\beta$ as $x \rightarrow \infty$,
which ensures that the magnetization has the correct form in the thermodynamic limit:
$\langle \lvert M \rvert ^p \rangle_{\infty} =  (-t)^{p\beta} \theta(-t).$
One can define a family of quotient functions~\cite{Binder81}
\begin{equation} \label{EQ:Binder_ratio}
Q_p(T,L) = \frac{\langle M^{2p} \rangle_L}{\langle M^2 \rangle_L^p} 
= \mathcal{N}(L)q_p(tL^{1/\nu}-\epsilon(L))
\end{equation}
---let us call them \emph{Binder ratios}---constructed in such a way that the leading-order $L$ dependence outside the scaling function factors out. 
The temperature dependence appears only in the argument of the function 
$q_p(x) = g_{M^{2p}}(x)/[g_{M^2}(x)]^p$. 
The subleading corrections enter as shown on the right-hand side of Eq.~\eqref{EQ:Binder_ratio}. 
The two lowest order Binder cumulants [Eqs.~(11) and (12) of Ref.~\onlinecite{Binder81}] correspond to the linear combinations $U = 1 - \tfrac{1}{3}Q_2$ and $V = 1 - \tfrac{1}{2}Q_2 + \tfrac{1}{30}Q_3$. 

In the thermodynamic limit, the Binder ratios jump discontinuously between
$Q_p(T<\crit{T}) = q_p(-\infty) = 1$ and $Q_p(T>\crit{T}) = q_p(\infty)$. The value
of $q_p(\infty) \neq 1$ is simply a prefactor of the gaussian distribution
of thermally randomize spins; the value $Q_p(\crit{T}) = q_p(0)$ is a non-trivial, universal constant. 
The Binder cumulants are also step functions, taking the values $U = 2/3$ and $V = 8/15$ 
in the ordered phase and $U=V=0$ in the disordered phase.

We now consider the point of intersection $\cross{T}(L,L')$ between two Binder ratio curves for system sizes
$L \neq L'$.
To start, let us assume that $\cross{T}(L,nL)$, for some fixed ratio $n = L'/L$, converges to $\crit{T}$ as a function of $L$ \emph{faster} than $L^{-1/\nu}$. In that case, $\cross{t}(L,nL)L^{1/\nu}$ is a small quantity. 
Accordingly,
\begin{equation}
Q(\cross{T},L) = \mathcal{N}(L)\Bigl[q_0 + q_1\bigl(\cross{t}L^{1/\nu}-\epsilon(L)\bigr) + \cdots \Bigr].
\end{equation}
[We have dropped the $p$ label; the subscripts here refer to the expansion coefficients
of $q(x)$ about $x=0$.]
Hence, the crossing criterion $Q(\cross{T},L)=Q(\cross{T},L')$ 
implies that 
\begin{equation} \label{EQ:TiLnLunsimplified}
\cross{t} = \frac{-q_0\bigl[\mathcal{N}(L)-\mathcal{N}(L')\bigr]
+q_1\bigl[\mathcal{N}(L)\epsilon(L)-\mathcal{N}(L')\epsilon(L')\bigr]}
{q_1\bigl[\mathcal{N}(L)L^{1/\nu}-\mathcal{N}(L')(L')^{1/\nu}\bigr]}
\end{equation}
or, in the notation of Eq.~\eqref{EQ:FSSsubleading},
\begin{multline} \label{EQ:TiLnL}
\cross{t}(L,nL) = \frac{cq_0}{q_1}\biggl(\frac{1-n^{-\omega}}{n^{1/\nu}-1}\biggr)L^{-1/\nu-\omega}\\
- d\biggl(\frac{1-n^{-\phi/\nu}}{n^{1/\nu}-1}\biggr)L^{-(1+\phi)/\nu}.
\end{multline}
[In retrospect, the assumption that $\cross{T}(L,nL) \rightarrow \crit{T}$ faster than $L^{-1/\nu}$ 
was justified.]
Note that although Eq.~\eqref{EQ:TiLnL} was derived with the Binder ratio in mind,
it applies equally to any quantity $A(T,L)L^{-\kappa/\nu}$ obeying Eq.~\eqref{EQ:FSSsubleading} 
whose scaling function is bounded and monotonic.

\begin{figure*}
\includegraphics{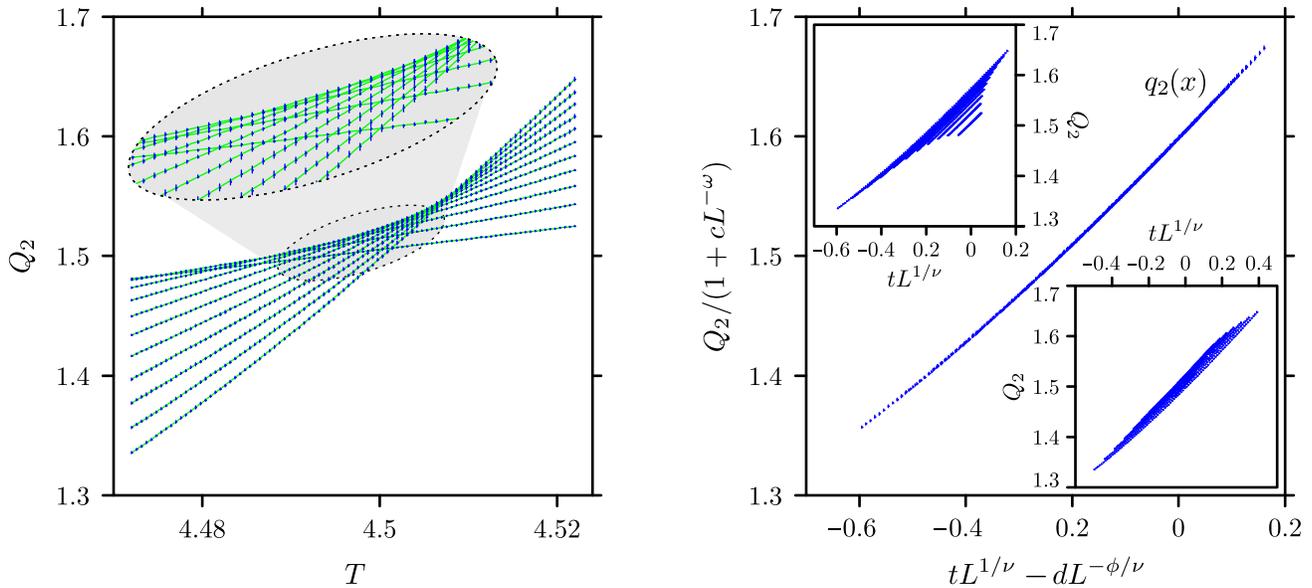}
\caption{\label{FIG:ising3d} 
(color online)
The second Binder ratio $Q_2(T,L)$, computed via Monte Carlo for the 3D Ising model on lattices of size $L = 4,5,\ldots,14$. In the left-hand panel, the solid (green) lines are guides to the eye, connecting data points computed for equal system sizes; the curves grow steeper with increasing $L$. The oval inset presents a magnified view in which the individual data points 
and their error bars (blue) are more clearly visible. 
The scaling plot in the right-hand panel shows the complete collapse of the data onto a single curve.
This best fit gives 
$\crit{T} = 4.5114(2)$, $\nu = 0.625(1)$, and 
$q_2(0) = 1.60(1)$ [or $U^* = 1-\frac{1}{3}q_2(0)$ = 0.467(3)].
The two inset panels are purely leading-order scaling plots ($c=d=0$) based on (top-left)
the best fit values just quoted and (bottom-right) the values $\crit{T} = 4.500(2)$, $\nu = 0.60(1)$  obtained by fitting  $Q_2$ versus $tL^{1/\nu}$.
}
\end{figure*}

\emph{Numerical results}---We have applied these results 
to three numerical test cases: the two- (2D) and three-dimensional (3D) nearest-neighbour Ising models,
and the double-layer quantum Heisenberg antiferromagnet~\cite{Shevchenko00}.
Monte Carlo data were generated for the Ising models using
the Swendsen-Wang cluster update algorithm~\cite{Swendsen87}
and for the quantum antiferromagnet using stochastic series expansion~\cite{Sandvik99}.

The Ising models exhibit classical, thermally-driven phase transitions
at $T_{\text{c,2D}} = 2/\log(1+\sqrt{2}) \approx 2.2691853$ 
and $T_{\text{c,3D}} \approx 4.511$.
The antiferromagnetic bilayer, on the other hand, exhibits a zero-temperature, quantum
phase transition as the interlayer coupling strength $\alpha$ is tuned through
its critical value $\crit{\alpha} \approx 2.5218$~\cite{Shevchenko00,Wang05}. 
For each of these cases, we computed the second Binder ratio over
a fine, uniform mesh of temperature (coupling) values in a fixed interval 
containing $\crit{T}$ ($\crit{\alpha}$). 
(In practice, when rough estimates of $\crit{T}$ and $\nu$ are available in advance, 
it makes more sense to take measurements in 
the range $\lvert x \rvert = \lvert t \rvert L^{1/\nu} \lesssim 1$.)
The simulations were performed 
for a range of relatively small lattice sizes.

The data were fit to Eq.~\eqref{EQ:FSSsubleading} (with $\kappa = 0$) 
using the the Levenberg-Marquardt nonlinear optimization algorithm~\cite{Gill81},
which searched the space of parameters
$\crit{T}$ ($\crit{\alpha}$), $\nu$, $\omega$, $\phi$, $c$, and $d$
to produce the best collapse of the data.
Uncertainties in those values were
computed by bootstrapping~\cite{Efron93} the regression over the original Monte Carlo data.
Comparison with the 2D Ising model, where the solution is known~\cite{Onsager44}, 
serves as an important proof of concept: the fitted values $\crit{T} = 2.26917(2)$ and $\nu = 0.99(1)$ 
with $L = 4,5,\ldots,32$ agree within statistical uncertainties with the exact results. 
The same procedure applied to the quantum bilayer 
with $L = 8,10,12,\ldots,42$
yields $\crit{\alpha} = 2.52181(4)$, the most accurate value to date,
and $\nu = 0.715(2)$, which is consistent with the
3D classical Heisenberg universality class, as expected~\cite{Chakravarty88}.

The raw data for the 3D Ising model are shown in the left-hand panel of Fig.~\ref{FIG:ising3d}.
In the right-hand panel, the same data are rescaled according to Eq.~\eqref{EQ:FSSsubleading}
and collapse convincingly onto a single curve.
The two inset panels on the right illustrate the failure of naive FSS:
plotted in the conventional reduced coordinates ($c=d=0$), the data are clearly distinguishable
as a series of separate curves corresponding to different lattice sizes.

The fitted value of the critical temperature $\crit{T} = 4.5114(2)$ is
consistent with the Rosengren conjecture~\cite{Rosengren86}
and with other reliable estimates; \emph{cf.} Ref.~\onlinecite{Talapov96} and
Table 19 in Ref.~\onlinecite{Blote95}.
Note that our result is nearly as accurate as those of
Pawley \emph{et al}.\ and Garcia {\emph{et al}.\
(\emph{viz.}, $\crit{T} = 4.5115(1)$ in Ref.~\onlinecite{Pawley84}
and $\crit{T} = 4.51152(12)$ in Ref.~\onlinecite{Garcia03}),
which are based on simulations up to sizes $L=64$ and $L=115$, respectively.
And while our value of $\nu = 0.625(1)$ is small with respect to some 
estimates~\cite{Pawley84,Ferrenberg91,Blote95}, it appears to be in very good agreement with more 
recent Monte Carlo ($L = 90,100,115$)~\cite{Garcia03} and Monte Carlo Renormalization 
Group ($L = 64, 128, 256$)~\cite{Baillie92,Gupta96} calculations.
As an additional check, we can read off the value $q_2(0) = 1.60(1)$
(the ``y-intercept'' in the right-hand panel of Fig.~\ref{FIG:ising3d}),
which agrees with the value $q_2(0) = 1.604(1)$ computed by 
independent methods~\cite{Blote95}.

Figure~\ref{FIG:crossing} plots the Binder ratio intersection points 
of the 2D Ising model and the quantum bilayer for several system size ratios $n = L'/L$.
In each case, the complete set of crossing points was computed by 
interpolating smooth curves between the measured $Q_2$ values
(as in the left panel of Fig.~\ref{FIG:ising3d}, where the grid lines form
a set of intersection points). The data were successfully fit to the scaling form
given by Eqs.~\eqref{EQ:TiLnLunsimplified} and \eqref{EQ:TiLnL}.
The resulting estimates $\crit{T} = 2.2692(8)$, $\nu = 0.99(2)$ 
and $\crit{\alpha} = 2.5218(2)$, $\nu = 0.715(8)$ are somewhat less accurate than
but consistent with the values determined by the data collapse analysis.

\begin{figure}
\includegraphics{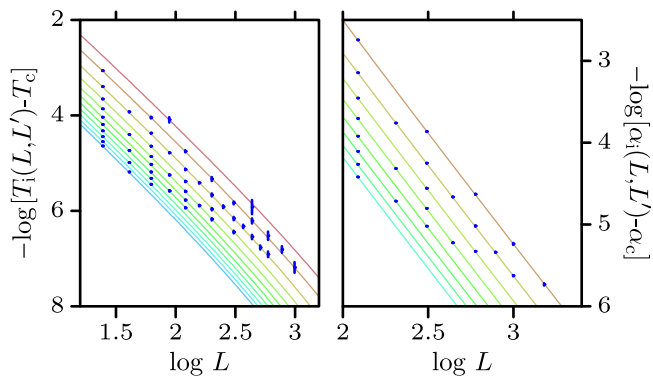}
\caption{\label{FIG:crossing}
(color online)
Binder ratio crossing points $\cross{T}(L,L')$ plotted
with respect to $L$ ($< L'$) for the (left) 2D Ising model
and (right) double-layer quantum antiferromagnet.
The solid lines depict the analytic result given in 
Eq.~\eqref{EQ:TiLnL} for lines of
constant $n = L'/L$.
From top (red) to bottom (blue), 
the ratios shown are (left) 
$n = 8/7$, $n = 3/2, 2, \ldots, 11/2, 6$ (by increments of $1/2$)
and (right) $n = 3/2, 2, \ldots, 9/2, 5$.
}
\end{figure}

\emph{Conclusions}---Finite-size scaling describes how critical behaviour emerges from finite 
systems in the limit of increasingly large system size. As $L\rightarrow \infty$, 
quantities in the critical region become 
universal modulo a size-dependent rescaling of the variables, a property which, in principle, 
can be exploited to measure unknown critical parameters by means of data collapse.
In practice, such an analysis is likely to give misleading results.
Still, we have shown that data collapse can be made a viable, high-accuracy analysis tool,
so long as subleading corrections to finite-size scaling are properly taken into account. 
We have introduced a new way of expressing those corrections, in which two kinds of deformation---the renormalization $g_A \rightarrow \mathcal{N} g_A$ and the shift $g_A(x) \rightarrow g_A(x-\epsilon)$---serve as the fundamental deviations from leading-order FSS. The greater expressiveness of the scaling form typically leads to larger but more meaningful and reliable statistical errors on the 
critical parameters.

Large simulations devour computing resources. 
With Monte Carlo, CPU time scales as $L^{d+z}$, where $z$ is a characteristic dynamical exponent.
For the Ising model in $d=3$, Swendsen-Wang updates~\cite{Swendsen87} give $z \approx 0.75$.
This implies that it takes as much time to compute the
\emph{single} system size $L = 16$ as it does to compute \emph{all} of $L = 4,5,\ldots,12$.
(Even worse, for models where specialized update schemes are not available, $z\approx 2$.)
The success of Eq.~\eqref{EQ:FSSsubleading} as a fitting form 
well away from the scaling limit has an interesting implication:
it may be more profitable to collect a large quantity of easily-obtainable intermediate-$L$ data~\cite{endnote2} (so as to maximize the fit statistics)
than to make a herculean effort to obtain data for the very largest possible lattice sizes.


\begin{thebibliography}{99}

\bibitem{Fisher71} M.\ E.\ Fisher, in {\it Critical Phenomena}, Proceedings of the
Enrico Fermi International School of Physics, 
ed.\ M.\ S.\ Green (Academic, New York, 1971), Vol. 51.
\bibitem{Fisher72} M.\ E.\ Fisher and M.\ N.\ Barber, Phys.\ Rev.\ Lett., {\bf 28}, 1516 (1972).
\bibitem{Brezin82} E.\ Br\'{e}zin, J.\ Phys.\ (Paris) {\bf 43}, 15 (1982).
\bibitem{Barber83} M.\! N.\! Barber, in {\it Phase Transitions and Critical Phenomena}, ed.\ 
C.\! Domb (Academic, New York, 1983), Vol. 8.
\bibitem{Brezin85} E.\ Br\'{e}zin and J.\ Zinn-Justin, Nucl. Phys. B {\bf 257}, 867 (1985).
\bibitem{Stanley99} H.\ E.\ Stanley, Rev. Mod. Phys. {\bf 71}, S358 (1999).
\bibitem{Binder70_80} K.\ Binder, Thin Solid Films {\bf 20}, 367 (1974); D.\ P.\ Landau, Phys. Rev. B {\bf 13}, 2997 (1976); K.\ Binder and D.\ P.\ Landau, {\it ibid}. {\bf 21}, 1941 (1980).
\bibitem{Ferron04} A.\ Ferr\'{o}n and P. Serra, J.\ Chem.\ Phys.\ {\bf 120}, 8412 (2004).
\bibitem{endnote1} A suitable goodness of fit measure for data collapse
can be defined even with no knowledge of $g_A(x)$;
see S.\ M.\ Bhattacharjee and F.\ Seno, J. Phys. A {\bf 34}, 6375 (2001).
\bibitem{Binder81} K.\ Binder, Z.\ Phys.\ B {\bf 43}, 119-140 (1981).
\bibitem{Wang05} L.\ Wang, K.\ S.\ D.\ Beach, A.\ W.\ Sandvik, to be published.
\bibitem{Shevchenko00}P.\ V.\ Shevchenko, A.\ W.\ Sandvik, and O.\ P.\ Sushkov,
Phys. Rev. B {\bf 61}, 3475 (2000).
\bibitem{Swendsen87}R.\ H.\ Swendsen and J.-S.\ Wang, Phys. Rev. Lett. {\bf 58}, 86 (1987).
\bibitem{Sandvik99} A.\ Sandvik, Phys. Rev. B, {\bf 59}, R14157 (1999);
{\it ibid.} {\bf 56}, 11678 (1997).
\bibitem{Gill81}P.\ R.\ Gill, W.\ Murray, and M.\ H.\ Wright,  {\it Practical Optimization} (Academic Press, San Diego, 1988).
\bibitem{Efron93} B.\ Efron and R.\ Tibshirani, {\it An Introduction to the Bootstrap} (Chapman \& Hall/CRC, Boca Raton, 1993).
\bibitem{Onsager44}L.\ Onsager, Phys.\ Rev.\ {\bf 65}, 117 (1944).
\bibitem{Chakravarty88}S. Chakravarty, B.\ I.\ Halperin, and D.\ R.\ Nelson, 
Phys.\ Rev.\ Lett.\ {\bf 60}, 1057 (1988).
\bibitem{Rosengren86} A.\ Rosengren, J.\ Phys.\ A {\bf 19}, 1709 (1986).
\bibitem{Talapov96} A. L.\ Talapov and H.\ W.\ J.\ Bl\"{o}te, J.\ Phys.\ A {\bf 29}, 5727 (1996); {\tt cond-mat/9603013}
\bibitem{Blote95}H.\ W.\ J.\ Bl\"{o}te, E.\ Luijten, and J.\ R.\ Herniga, J. Phys. A {\bf 28}, 6289 (1995).
\bibitem{Pawley84} G.\ S.\ Pawley, R.\ H.\ Swendsen, D.\ J.\ Wallace, K.\ G.\ Wilson, Phys.\ Rev.\ B {\bf 29}, 4030 (1984).
\bibitem{Garcia03} J.\ Garcia and J.\ A.\  Gonzalo, Physica A {\bf 326}, 464 (2003); {\tt cond-mat/0211270}
\bibitem{Ferrenberg91}A.\ M.\ Ferrenberg and D.\ P.\ Landau, Phys. Rev. B {\bf 44}, 5081 (1991).
\bibitem{Baillie92} C.\ F.\ Baillie, R.\ Gupta, K.\ A.\ Hawick, and G.\ S.\ Pawley, Phys.\ Rev.\ B {\bf 45}, 10438 (1992).
\bibitem{Gupta96} R.\ Gupta and P. Tamayo, Int.\ J.\ Mod.\ Phys.\ C {\bf 7}, 305 (1996); {\tt cond-mat/9601048}
\bibitem{endnote2}Some care must be taken. 
There may be a crossover scale $\xi_{\times}$ such 
that $L < \xi_{\times}$ corresponds to an entirely different universality class.

\end{thebibliography}
\end{document}